\documentclass[10pt,a4paper]{article}

\usepackage{amsmath,amssymb,amsthm}
\usepackage{url}
\usepackage{placeins}
\usepackage{subfigure}
\usepackage{multirow}
\usepackage{epsfig}
\usepackage{chicago}

\newcommand{\bR}{\ensuremath{\mathbb{R}}}

\newcommand{\cD}{\ensuremath{\mathcal{D}}}
\newcommand{\cT}{\ensuremath{\mathcal{T}}}

\newcommand\be{$$}
\newcommand\ee{$$}
\newcommand\ben{\begin{equation}}
\newcommand\een{\end{equation}}
\newcommand\bea{\begin{eqnarray*}}
\newcommand\eea{\end{eqnarray*}}
\newcommand\bean{\begin{eqnarray}}
\newcommand\eean{\end{eqnarray}}

\newcommand{\Va}{\ensuremath{{\text{VAR}(\alpha)}}}
\newcommand{\CDF}{\ensuremath{{\text{CDF}}}}
\newcommand{\PDF}{\ensuremath{{\text{PDF}}}}
\newcommand{\ESa}{\ensuremath{{\text{ES}(\alpha)}}}
\newcommand{\ESb}{\ensuremath{{\text{ES}(\beta)}}}

\begin{document}
\author{Chris Kenyon\footnote{Contact: chris.kenyon@lloydsbanking.com} and Andrew Green\footnote{Contact: andrew.green2@lloydsbanking.com}}
\title{VAR and ES/CVAR Dependence\\ on data cleaning and Data Models:\\ Analysis and Resolution\footnote{\bf The views expressed are those of the author(s) only, no other representation should be attributed.}
}
\date{\today\\ \vskip5mm Version 1.01}

\maketitle

\begin{abstract}
Historical (Stressed-) Value-at-Risk ((S)VAR), and Expected Shortfall (ES), are widely used risk measures in regulatory capital and Initial Margin, i.e. funding, computations.  However, whilst the definitions of VAR and ES are unambiguous, they depend on input distributions that are data-cleaning- and Data-Model-dependent.  We quantify the scale of these effects from USD CDS (2004--2014), and from USD interest rates (1989--2014, single-curve setup before 2004, multi-curve setup after 2004), and make two standardisation proposals: for data; and for Data-Models.  VAR and ES are required for lifetime portfolio calculations, i.e. collateral calls, which cover a wide range of market states.  Hence we need standard, i.e. clean, complete, and common (i.e. identical for all banks), market data also covering this wide range of market states.  This data is historically incomplete and not clean hence data standardization is required.  Stressed VAR and ES require moving market movements during a past (usually not recent) window to current, and future, market states.  All choices (e.g. absolute difference, relative, relative scaled by some function of market states) implicitly define a Data Model for transformation of extreme market moves (recall that 99th percentiles are typical, and the behaviour of the rest is irrelevant).  Hence we propose standard Data Models.  These are necessary because different banks have different stress windows.  Where there is no data, or a requirement for simplicity, we propose standard lookup tables (one per window, etc.).  Without this standardization of data and Data Models we demonstrate that VAR and ES are complex derivatives of subjective choices.  
\end{abstract}
 
\section{Introduction}
Historical Value-at-Risk (VAR), and Expected Shortfall (ES, aka Conditional Value-at-Risk or CVAR) are required under current conditions, and under stressed market conditions, for capital and Initial Margin, i.e. funding, calculations \cite{BCBS-189,BCBS-261,BCBS-265}.  They are also required under future conditions for lifetime cost of funding, and lifetime cost of capital pricing \cite{Green2014a,Green2014b}.  Although the mathematical formulae for VAR and ES are unambiguous given an input distribution, we demonstrate that this input distribution is highly data cleaning and Data Model dependent.  Hence we make concrete standardisation proposals.

We quantify the scale of data cleaning and Data Model effects covering: daily USD senior unsecured CDS (2004--2014); daily USD interest rates using swaps (OIS and Libor) out to 30-year maturity (1989--2014); and daily Effective Fed Fund rates (1972--2014).  We use a single-curve approach before 2004 and a multi-curve (discounting and spread) approach after 2004 for interpreting the OIS and Libor swaps data \cite{Kenyon2012a,Morini2013a}.  The Effective Fed Fund rate data enables us to probe higher interest rate regimes (to 20\%) rather than only the last twenty years of lower rates.  Swaps data is unavailable much prior to 1989, and OIS swaps are unavailable much prior to 2004.  Understanding a wide range of rates is relevant for stressed VAR and stressed ES because these involve moving market movements from a past (usually not recent) window to current and future market states.   A wide range of states is also important for lifetime capital and funding calculations \cite{Green2014a,Green2014b} where simulation paths can disperse widely.  It is also relevant for understanding possible future market states before they occur.   

We also quantify the relative sensitivity of VAR and ES to data cleaning.  Roughly speaking, once data is clean we find no significant difference w.r.t. Data-Models between 10-day VAR(99\%) and ES(97.5\%) --- our cleaning removed outliers.  Both data-cleaning and Data-Models have significant, and separate, effects.  

Since VAR and ES are highly data cleaning and Data Model dependent we propose standard data and standard Data-Models.  Standard data is clean, complete and common, i.e. identical for all banks.  Where there is insufficient data, or a requirement for simplicity, we propose a standard look-up table approach.  This comprises sets of lookup tables that are appropriate for banks with different stress windows.  To avoid model risk we follow the prospective Prudential Valuation \cite{EBA-CP-2013-28} in proposing using a set of Data Models.  We also propose that standard Data Models and standard lookup tables are consistent with the historical record.  This may seem obvious but we will demonstrate that two common choices (absolute differences and relative differences) are not consistent with the historical record for USD interest rates 1989--2014.  Without standardisation of the input distribution construction, both VAR and ES are complex derivatives of subjective data cleaning choices and subjective Data Model choices.

\subsection{Terms}

The two key issues here for VAR and ES are data cleaning and data model which we define as follows.
\begin{itemize}
	\item Data Cleaning: the process data goes through before it reaches any analysis system.  The two key issues are missing data, and identification of false data (usually outliers), which can then be treated as missing data (assuming that the false data carries no relevant information).   An example of false data would be data entered incorrectly, e.g. 31 instead of 13.  Data received by an institution may have already been cleaned by a separate institution e.g. a commercial data provider.  Furthermore, data cleaning procedures may also change periodically making the issue more complex.
	\item Standard Data: data that is clean, complete, and common.  That is, it is identical for all banks.  This can be effected in the same manner that banks are subject to common capital regulations.
	\item Data Model: how historical time series data is turned into an input distribution for VAR and ES after data cleaning.  It includes how data is moved from a past historical window to current and future market states that may be very different.  A typical example would be the choice of whether to use relative differences, or absolute differences, or relative difference scaled by some function of market states, in 10-day VAR(99\%) for bilateral IM \cite{BCBS-261}.  In practice that are many choices, and some simple choices are not consistent with the historical record.  This is the second key issue investigated here.
	\item Standard Data Model: a data model that is identical for all banks.  This can be effected in the same manner that banks are subject to common capital regulations.
\end{itemize}

For completeness we provide standard definitions of VAR and ES here.  \Va\ is the lower bound on the loss that is expected to occur $\alpha$-percent of the time.  \ESa\ is the expectation of the loss, given that it is at least as large as \Va.  Standard definitions are:
\bea
\Va &:=& \min\{x \quad \text{s.t.} \quad \CDF(x)\ge \alpha\} \\
\ESb &:=& \frac{1}{1-\beta}\int_{\beta}^1 \Va d\alpha\qquad = \int_{\Va}^{\infty} x \PDF(x) dx
\eea
Where CDF is the Cumulative Distribution Function of the losses (this exists for all distributions), and PDF is the Probability Distribution Function of the losses.

\subsection{Previous Work\label{s:prev}}

Sensitivity of ES to outliers and estimation of the input distribution has been noted \cite{Cont2008a,BCBS-265}, but not the dependence on data cleaning and Data Model (e.g. relative or absolute differences).  Portfolio optimization under VAR and ES is know to be sensitive to noise \cite{Kondor2007a,Lima2011a}.  Parameter uncertainty has a long history.   However, data cleaning and Data Model issues are complementary but orthogonal to usual statistical questions such as estimation error \cite{Kondor2014a}.  

Data cleaning differences between banks, and differences in Data Models may have contributed to the range of outcomes observed in comparative risk weighted assets for market risk \cite{BCBS-240}, as that document also remarked.  

There are many studies of implied volatility under risk-neutral and historical measures, \cite{Rebonato2004a,Hull2014b,DeGuillaume2013a}.   Two of the most recent find a three-part behaviour of volatility versus level (also considering different tenors): increasing Normal volatility (of 1-day differences) with level up to around 2\%\ then constant Normal volatility up to about 6\%\ then increasing Normal volatility again up to around 10\%.  These used data from 2004--2010 in one instance, and starting much further back  in the other.  The longer study had to use bond data to go back and so mixing funded (bond) and unfunded (swap) data.  Both studies finishing before the current low rates regime (roughly 2010--onwards) had lasted very long.  We show roughly similar results when considering Normal volatility but find a cleaner signal for log-Normal volatility.  We also demonstrate issues with mixing funded and unfunded data.  Our results also cover discounting and spread (i.e. multi-curve) behaviour, after 2004, \cite{Kenyon2012a,Morini2013a}.  We show that spread curves (difference between projection and discount curves) behave differently to discount curves.   Our results are robust with respect to tenor, so this represents an addition to earlier work with the benefit of additional data and methods.  

The contributions of this paper are: firstly defining the data cleaning (missing and false data) and Data Model issues; secondly quantifying their (different) effects for both VAR and ES; identifying and quantifying the issues for stressed VAR and ES to do with applying past window data to current (and future) market conditions; thirdly standardisation proposals for standard market data (clean, complete, and common) and standard Data Models.   We also contribute a lookup table alternative where there is either insufficient data or a requirement for simplicity.  
  
\section{VAR and ES Regulations\label{s:reg}}

Here, in Table \ref{t:targets}, we detail places that VAR and ES must be used now, or are proposed, for regulatory purposes.  In BCBS-265 (FRTB) VAR(99\%) is replaced for requirements by ES(97.5\%), but VAR(99\%) is retained for backtesting. For IMM cases, CVA VAR capital and Market Risk (MR) capital use the sum of the amounts derived from VAR and SVAR.  (CCR uses the maximum, but is not derived from VAR or SVAR).  Using the sum of VAR and SVAR derived capital for MR and CVA is obviously pro-cyclical because VAR is pro-cyclical.  Central Counterparties (CCPs) may also use VAR or ES type methodologies.  For example LCH uses an ES-type methodology, but details are proprietary and subject to change.
\begin{table}[htbp]
		\centering
			\begin{tabular}{ccp{1.6cm}p{1.6cm}p{1.2cm} p{1.4cm}ccc}
			\multirow{2}{*}{Metric} & \multirow{2}{*}{Use} & \multirow{2}{*}{Source}    & \multicolumn{6}{c}{Definition} \\ \cline{4-9}
			          &     &           & holding period & window length   & stress?& obs &  \%-ile & in tail \\ \hline				
				VAR     & MR  & BCBS-128  &   5/10/20$\dag$ &  recent 1Y     & NA        &    260      & 99 & 2.6   \\
				SVAR    & MR  & BCBS-158  &   5/10/20$\dag$ & 1 in 3Y        & Y single  &    260      & 99 & 2.6   \\
				VAR     & CVA & BCBS-189  &    10           & recent 1Y      & NA        &    260      & 99 & 2.6   \\
				SVAR    & CVA & BCBS-189  &    10           & 1 in 3Y        & Y single  &    260      & 99 & 2.6   \\
				VAR     & IM  & BCBS-261  &    10           & 1--5Y          & Y by AC   &  260--1300  & 99 & 2.6--13  \\
				VAR     & IM  & Art.11(15) draft &    10    & recent $\ge$3Y & $\ge$25\%\ by AC & 780-- & 99 & 7.8-- \\
				ES      & All & BCBS-265  &   5/10/20/1Y    & 1Y etc.        & varies    &  260--      & 97.5 & 6.5--
				\\ \hline
			\end{tabular}
	\caption{Regulatory uses of VAR, and ES equivalent (last row).  Percentiles are one-sided, and VAR calibrations must be updated at least quarterly (or for Art.11(15) every six months).  Y = Yes; AC = Asset Class; NA = Not Applicable.  $\dag$  Depends on use case (repo / usual / secured lending), and can be increased if there are collateral disputes.  Generally the holding period given is the minimum which assumes daily remargining where appropriate.  Holding period can be increased because of collateral disputes in some cases.}
	\label{t:targets}
\end{table}

Art.11(15) in Table \ref{t:targets} refers to draft regulatory technical standards \cite{EBA-OTC-CCP2014a}.  We also remark the very small number of points upon which historical VAR or ES depend, roughly from two or three to ten points.

Note that tail risk metrics do not depend on non-tail data {\it therefore the behaviour of most of the data is irrelevant --- only the outliers count}.  We shall see later how this influences our analysis.  To anticipate, this makes us more interested in methods that are sensitive to outliers and less interested in statistically robust methods.  We have the opposite requirement from usual since we want to be tail-sensitive.

\section{Data Cleaning}

Data cleaning covers false data identification and missing data construction.  In this section we quantify the scope of missing data in one major asset class and market (USD senior unsecured CDS, i.e. USD-SU-CDS) where the extent of missing data is perhaps the most significant for regulatory capital (i.e. CVA VAR).  We look at both bulk USD-SU-CDS and GIPPS Euro-Sovereign CDS because there was both a financial crisis in 2008-9 and a Euro-Sovereign crisis in 2011-13.   The objective of this paper is not to propose any particular false data identification method or missing data construction algorithm.  Anything we proposed, however good, would only have a partial take-up amongst banks.  Instead we make the case for standardized data and standardized Data Models --- for regulatory purposes --- by quantifying the scale of the problem open to subjective solutions.  The next section on Data Models complements this section where we quantify missing data {\it effects} by quantifying the effects of data cleaning on VAR and ES and the effects of different Data Models as well.

The quantitative effects of false data identification and missing data construction has previously been neglected for regulatory VAR and ES, and especially with respect to stress windows. This is despite the fact that they are highly significant for tail metrics and that stressed VAR is usuaally by far the larger of VAR-derived capital requirements.  We quantify the effects of false data identification and missing data construction for USD interest rate data in the next section on Data Models (we require a Data Model before we can calculate an effect).

The effect of noise for VAR- and ES-based portfolio optimization has been observed \cite{Kondor2007a,Lima2011a}.  Since they are so significant we may ask whether we should prefer a model-based approach over historical data.  The extreme values reached during the 2008--onwards crisis have pushed attention towards historical approaches (e.g. in \cite{BCBS-261}, because model-based approaches may have been perceived to have failed.  Indeed the model governing tail events may not be the same as the model governing non-tail events so a model-based approach is not an automatic solution.  Assuming that we do use an historical approach we must be aware of its limitations, and deal with them: this is the subject of the current paper.

\subsection{Data Cleaning Governance}

Obviously there should only be one source of clean data in a bank, and the cleaning algorithms should be governed in the same way as pricing algorithms.  For capital and funding VAR and ES are part of derivative manufacturing costs specified in regulations and so feed directly in to pricing.

\begin{figure}
	\centering
		\includegraphics[width=0.70\textwidth]{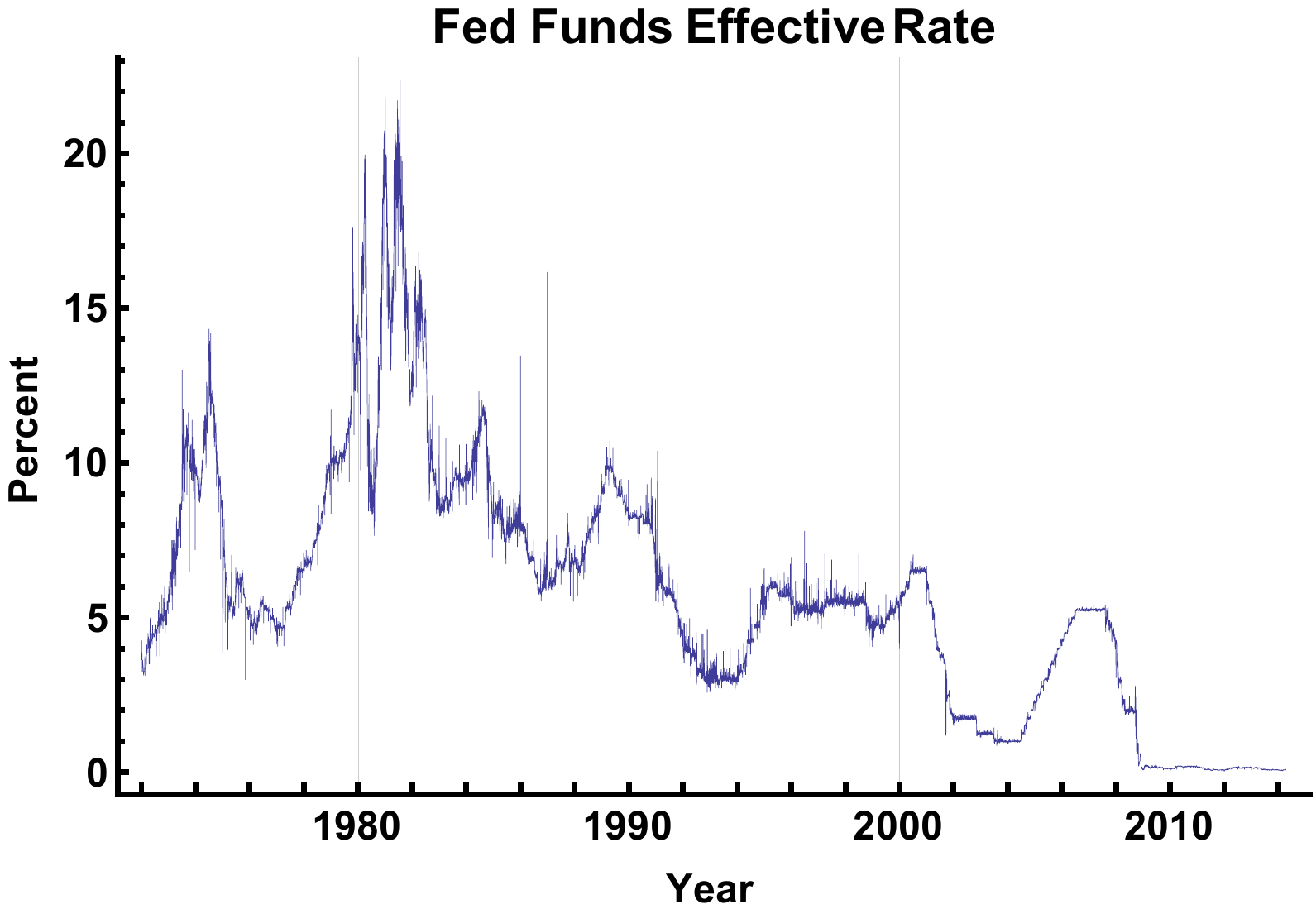}
	\caption{Daily Fed Funds Effective Rate from January 1972 to present from a major data provider.  Note that this data has not been altered at all after download.  There appears to be a change in quality (different fuzziness) around the turn of the millennium.}
	\label{f:cleaningDirty}
\end{figure}

If data is not clean, or cleaning is not subject to governance, the odds are high that VAR and ES results have major artefacts subject to arbitrary revision. As a complicating factor, commercial data providers update their data cleaning methods from time to time.  Consider Figure \ref{f:cleaningDirty} showing the  Effective Fed Funds Rate daily.  Using single high or low points as an indicator of clean data it is apparent that there was a change around the year 2000.  Given that this was the turn of the millennium, it could be that all the data participants improved their systems, or that the data provider updated their cleaning algorithms.  We will also use this dataset in the next section on Data Models.

\subsection{USD Senior Unsecured CDS: Missing Data}

In this section we quantify the extent of the missing data issue for CVA VAR for one important section of the CDS universe, USD senior unsecured (SU) CDS.  We consider both bulk CDS (all names) and sovereign (GIIPS) CDS. We pick credit because the missing data issue is most significant for this asset class amongst the regulatory asset classes (Rates, FX, Credit, and Commodities) in a major market (USD).   We used CDS data from a major CDS supplier and a report from that supplier where the most liquid CDS for each entity was given.  Hence there are different document clauses in the data used, but this is not part of the analysis.  

One motivation for the introduction of CVA VAR capital was that 2/3 of losses over the crisis were credit losses that were not defaults \cite{BCBS-189}.  The data behind this statement has not been made available but our analysis will shed some light on the accuracy limits of the statement.

\begin{figure}
	\centering
		\includegraphics[width=0.99\textwidth]{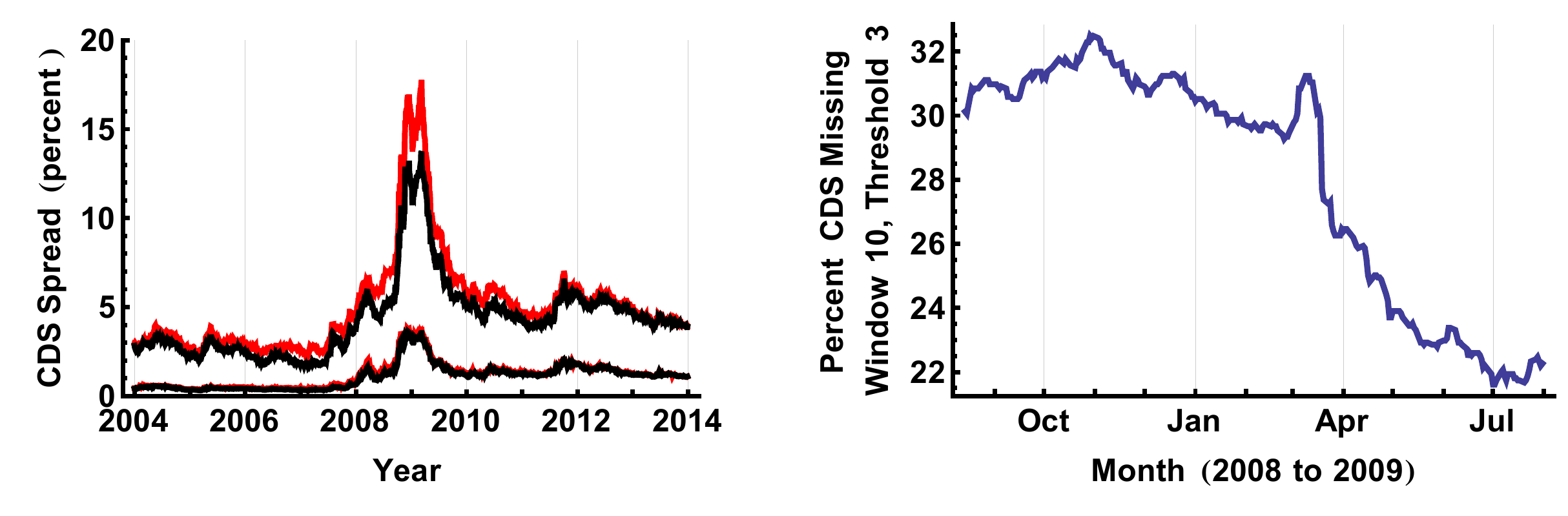}
		
		\includegraphics[width=0.55\textwidth,trim=0 0 0 0]{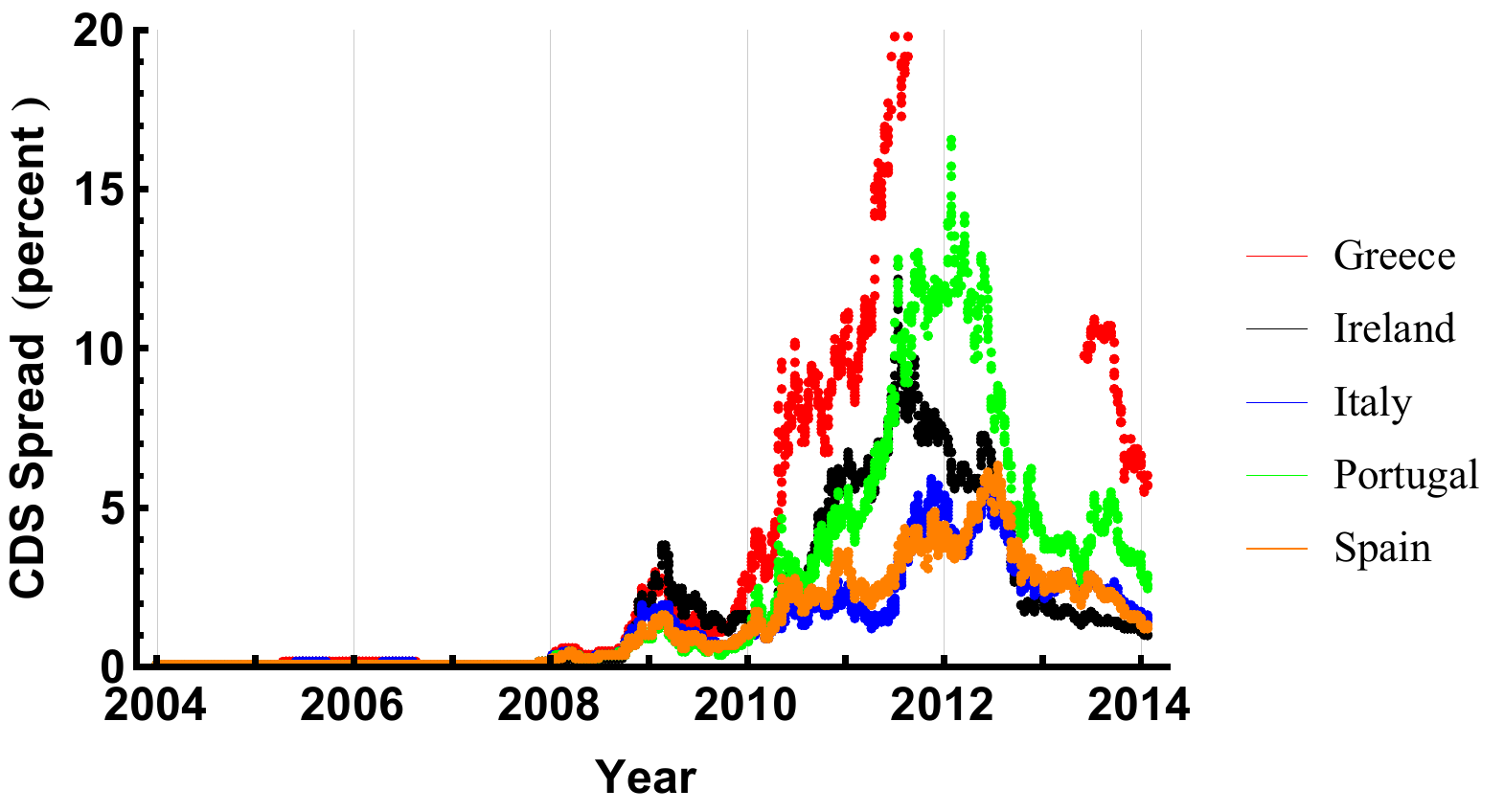}  \quad
		\begin{tabular}[b]{lp{1.8cm}}
		Sovereign & \%\ Missing \\ \hline
		Greece &  13 \\
		Ireland & 11 \\
		Italy & 0 \\
		Portugal & 0 \\
		Spain & 0 \\
		& \vphantom{V} \\
				& \vphantom{V}
		\end{tabular}
	\caption{{\bf TOP Left } shows the median (lower lines) and 90th percentile (upper lines) CDS spread of the USD-SU universe for ten years after 2004.  Red lines cover the universe, black lines the CDS that were available on Jan 1st 2014.  Peak market credit stress is from roughly Sept 2008 to Aug 2009.  {\bf TOP Right } shows the percentage of CDS that were available on Jan 1st 2014 that have at least three missing data points in a 10 business day window over the peak stress period.
	{\bf BOTTOM Left } GIIPS sovereign CDS, each daily observation is one dot.  Note that the vertical axis is limited to 20\%\ for ease of comparison --- Greek spreads went far higher. {\bf BOTTOM Right }  Percent of data missing 2004--2014.  During the crisis period 2011--2013 only Greece had missing data (42\%).
	}
	\label{f:CDSmissing}
\end{figure}

To identify appropriate stress periods, Figure \ref{f:CDSmissing} shows median and 90th percentile CDS spreads over the USD-SU universe (top row), and GIIPS (Greece, Italy, Ireland, Portugal, Spain) sovereign CDS data.  We call the first stress period ''Financial'' and the second stress period ``Euro-Sovereign''.  The higher red lines show the whole universe, whilst the lower black lines show the values for the CDS that were available on Jan 1st 2014 (i.e. currently).  Roughly 20\%\ of the CDS names currently available are not available at all over the Financial stress period.  During the height of the stress roughly 30\%\ of CDS names available currently available have gaps that cover the entirety of regulatory VAR or ES tail requirements (three points).  CDS names here correspond to legally identifiable entities.  During the Euro-Sovereign stress period only Greek CDS were missing, but these were absent for 42\%\ of the period.

A US bank may have as their single-stress period 2008-9, whereas a Euro-focussed bank could well have their single-year single-stress period within 2011-13.

Bonds are not a substitute for CDS over the stress period, because the stresses combined liquidity (of different sorts, inter-bank, and Euro-Sovereign) as well as credit.  Credit and liquidity effects on bonds are difficult to separate.  Thus we have no suitable substitute.  Even if we did have a substitute for CDS spreads, the regulations \cite{BCBS-189} explicitly require the use of CDS spreads for CVA VAR.  Given that CDS provide explicit relief from CVA VAR capital (where not exempt, as in Euro-Sovereigns under CRD IV) it is natural to ask how this may have distorted their interpretation as market-implied default probabilities.  \cite{Kenyon2013d} finds that up to 50\%\ of the CDS spread may be a payment for capital relief rather than default protection.

This lack of data has been recognised qualitatively and the regulations permit the use of proxies based on Sector, Rating and Region.  As also previously pointed out \cite{Chourdakis2013a}, the CDS universe is insufficient to provide coverage using this approach.  Technical suggestions for alternate proxy construction are available, e.g. \cite{Chourdakis2013a}.  However, in all these cases we are essentially constructing a mapping which will lack the idiosyncratic risk and create mapping risk.  Given that 30\%\ of the CDS universe is missing for the stress period it is possible that significant systematic risks are also missing.

\begin{table}[htbp]
		\centering
			\begin{tabular}{cc}
			Legally Unique CDS Names & Number Available \\ \hline
			quoted at least once 2004--2014   &  2958 \\
			available 1st Jan 2014 &  1527 \\
			at some point in stress period Sept 2008-- Aug 2009 & 78\% \\
			throughout stress period & 68\%
			\end{tabular}
	\caption{Availability of USD Senior Unsecured CDS for Stressed VAR or Stressed ES.  The table specializes as it goes down, so the 70\%\ number refers to the 1527 names currently available, not to all the names that have ever been observed.}
	\label{t:missing}
\end{table}
Given the extent of the missing data, validation of reconstructed data may be problematic.  We term these risks mapping risk and idiosyncratic reconstruction risk.  The scale of the problem is much larger than we have indicated because most banks have tens of thousands of counterparties so most will not be not covered by currently available, or historically available, CDS.  

We conclude that the extent of the missing data in the CDS universe, summarized in Table \ref{t:missing} is such that there will be wide variation between institutions in their reconstruction of the missing data.  Next we turn to the quantitative effects of data cleaning, so we move to Data Models.

\section{Data Model}

By Data Model we mean how time-series data is transformed into an input distribution \cD\ for use in VAR and ES.  For pricing lifetime capital and lifetime funding we need the input distributions given simulated future market states as well as the current market state.  Three typical Data Models are described below as examples: absolute, relative, and level-relative.  Again we do not imagine that this encompasses the creativity of data modelers, it is simply sufficient to quantify the output range of a set of common choices.

We suppose that the originally observed daily time series is: $\{x_i: i=1,\ldots,x_n\}$.  $n$ will typically be around 260 for a year of observations.  We further suppose that the calculation interval is $m$-days, which is typically 10 days (although it can be as short as five or as long as 20 or longer in some cases, e.g. if there are disputes on collateral calls).  The elements $\Delta_i$ of the input distribution, \cD, for VAR and ES are then generated as follows.
\begin{description}
	\item[Absolute] 
	\be
	\Delta_i = x_i - x_{i-m},\qquad i=m+1,\ldots,n
	\ee
	\item[Relative] 
	\be
	\Delta_i = \frac{x_i - x_{i-m}}{x_{i-m}},\qquad i=m+1,\ldots,n
	\ee
	\item[Level-Relative] 
	\be
	\Delta_i = f\left(l_j,l_{\text{now}},\frac{x_i - x_{i-m}}{x_{i-m}}\right),\qquad i=m+1,\ldots,n
	\ee
	Where $l_j$ describes some state of the market calculated on data up to $t_j$ (which will be some date relevant to the observed daily time series), and $l_{\text{now}}$ describes some state of the market calculated on data up to now (i.e. the date of the VAR or ES calcualtion).  $f()$ is the transformation of the relative difference according to the two market states.  
	
	For Level-Relative $l_*$ will be a metric giving the level relevant to the relative difference.  In the case of a linear level function we have:
	\be
	f_{\text{linear level-relative}}(l_i,l_{\text{now}},y_i) = y_i \times \frac{b\times l_{\text{now}} +a}{b\times l_i+c}
	\ee
	Alternatively, for a quadratic level function we have:
	\be
	f_{\text{quadratic level-relative}}(l_i,l_{\text{now}},y_i) = y_i \times \frac{c\times l_{\text{now}}^2+b\times l_{\text{now}} +a}{c\times l_i^2+b\times l_i+a}
	\ee
	where $a,b,c$ are the coefficients of the linear and quadratic level functions, and $l_y$ is the level relevant for the observation $y_i$ at $t_i$.  See below for an example, e.g. Figure \ref{f:USDdiffs}.
\end{description}

In addition to normal use, without a data model we cannot quantify the effect of missing or false data.  We know that data is altered in the original time-series but we would not know what the effect would be on the input distribution \cD\ and hence on VAR and ES.

\subsection{Intention and Expression}

The Data Model combines two distinct aspects, intention and expression.
\begin{itemize}
	\item Intention of the transformation.  For example an input distribution \cD\ created for SVAR would aim to preserve the stress in the original observations.  
	\item Requirements to express the intention given market characteristics.  For example, if there is a natural finite scale for value changes.  In this case, as the market level decreases, relative volatility will increase for constant market stress.   Considering the SVAR example the aim would be to express the same amount of stress but conditioned on current market state.
\end{itemize}
Hence the Data Model can be embodied by a transformation function \cT;
\be
\cT: (\bR^{n_o},\bR^{n_c}) \times \bR^{n_u} \mapsto \cD({n_d})
\ee
That is:
\be
\cT: (\text{observations}(t_o),\text{market state}(t_c)) \times \text{market state}(t_u))  \mapsto \text{distribution}(t_u)
\ee
Where:

$n_o$ number of observations;

$n_c$ numbers characterizing the market state relevant for the observations;

$n_u$ numbers characterizing the market state relevant for use of the observations;

$n_d$ number of points in empirical (historical) distribution \cD;

$t_o$ observation start;

$t_c$ start time for market state relevant for observations;

$t_u$ start time for market state relevant for creation of the shock distribution, i.e. use time.

All transformations implicitly identify a set of intentions and expressions, including assumptions on how a given market behaves.  What we do here is make these intentions and expressions explicit.  For example, using absolute differences implicitly assumes that the market is level-independent.  Using relative differences assumes that the market has a linear relationship with level that has no offset (i.e. the linear relationship goes through the origin).  We note that many authors have studied physical-measure market dynamics (we described several in Section \ref{s:prev}), but as far as we are aware they have not applied their results to tail metrics.

Typical intentions include:
\begin{itemize}
	\item preserve the market state of the observations;
	\item preserve the current market state.
\end{itemize}
However, without knowing how a market behaves given constant state, e.g. not-stressed, it is not possible to specify the transformation \cT\ that will express these intentions effectively.  Notice also that for tail metrics, e.g. VAR(99\%) we are only interested in the transformation of one specific point (the 99th percentile).  The behaviour of the market in general, i.e. the majority of the data, is irrelevant.

Stressed markets typically depict a mixture of location and scale changes.  Consider the Effective Fed Funds rate in Figure \ref{f:cleaningDirty}.  The Oil Shock of 1972 was mostly depicted by a level increase, whereas the financial shock of 2008 involved a level decrease.  The inflation period of the mid 1980s involved both a level increase and a scale increase.  We will go into detail below.

We would like to be able to provide statistical validation of Data Models.  However, given the lack of data, e.g. no USD swaps market data prior to 1989, or the existence of only one period of low rates after that date, this is unrealistic.  We can provide assumptions which can help, e.g. different country's behaviours' are the same, but these assumptions themselves are difficult to validate.  This highlights again the subjective nature of Data Models.  

\subsection{Intention, Expression, and Data Cleaning}

As mentioned above, without a data model we cannot quantify the effect of missing or false data on VAR and ES.  Here we do a set of examples using the daily Effective Fed Funds data shown in Figure \ref{f:cleaningDirty}.  In the next section we combine this with a detailed quantitative analysis of daily USD interest rates bootstrapped from OIS swaps and Libor swaps.  The aim in this section is to illustrate, qualitatively, the interaction of Intention, Expression, and Data Cleaning.  We  assume that we are interested in building the input distribution \cD\ for the current date (i.e. at the end of the illustrated data).
\begin{itemize}
	\item Step 1: characterize the market.  We choose to characterize it in terms of level, and state-and-scale (of movements).
	\begin{itemize}
		\item  Level: we observe that the data ranges between roughly 0\%\ and 25\%, thus we may decide to reject any model which produces negative values in the shock distribution \cD\ when applied to the current market state.  \footnote{Alternatively we might decide on a fixed lower bound, e.g. $-1\%$ from some additional knowledge of the market characteristics, e.g. announcements from the Fed about their willingness to implement negative rates.}
		\item State-and-Scale.  We assume that the market has stressed and non-stressed periods (e.g. regime-switching).  
		
		After August 2009 we know from the USD CDS universe that there is relatively little systematic credit stress on the US market.   Thus at low ($<$2\%) rates levels we only have unstressed data.  Although CDS data do not go back to 1972 we know that the two peaks in Fed Funds were results of the Oil Shock and a period of high inflation, both stress scenarios.  Thus above roughly 10\%\ we only have stressed data.  
		
		For current purposes we assume that all other dates have a mix of stressed and unstressed observations, but with a major preponderance of unstressed observations.
	\end{itemize}
	\item Step 2: Define Intention.  Let us assume that we want to create an input distribution \cD\ for SVAR.  That is, we want to preserve the stress in any chosen observations.  
	\item Step 3: Characterize Expression. We must ask how stress is expressed at the current (low) market level.  Given that we have no stressed data at the current (low) market level, we can instead ask the opposite question: how is a non-stressed market expressed?
\end{itemize}

\subsection{USD Interest Rates Analysis}

We focus here on USD interest rate data from 1989 (start of extensive swaps data) through 2014, specifically OIS swaps, 3M deposits, and Libor swaps out to 30-year maturity.  For 1989--2004 we assume that the market priced using a single-curve appoach and for 2004--2014 we assume a multi-curve approach (discounting and spread), so that we can get a picture of both discount and spread behaviour.  The potential mixture of pricing approaches from 2004--2007, i.e. different banks were probably using different approaches, is not an issue because the observed spreads (OIS to Libor) were very low.   This data is not complete nor is it clean.  Appendix 1 describes the completion and cleaning procedure.  We include cleaning effects explicity via an analysis of Effective Fed Funds data for simplicity (because this involves a single time-series not a curve)  Although the Effective Fed Funds data might be expected to be high quality (i.e. very clean) it is not obviously so.

\paragraph{IRS Tenor-Point Volatility vs Level}

\begin{figure}[htbp]
	\centering
		\includegraphics[width=0.99\textwidth]{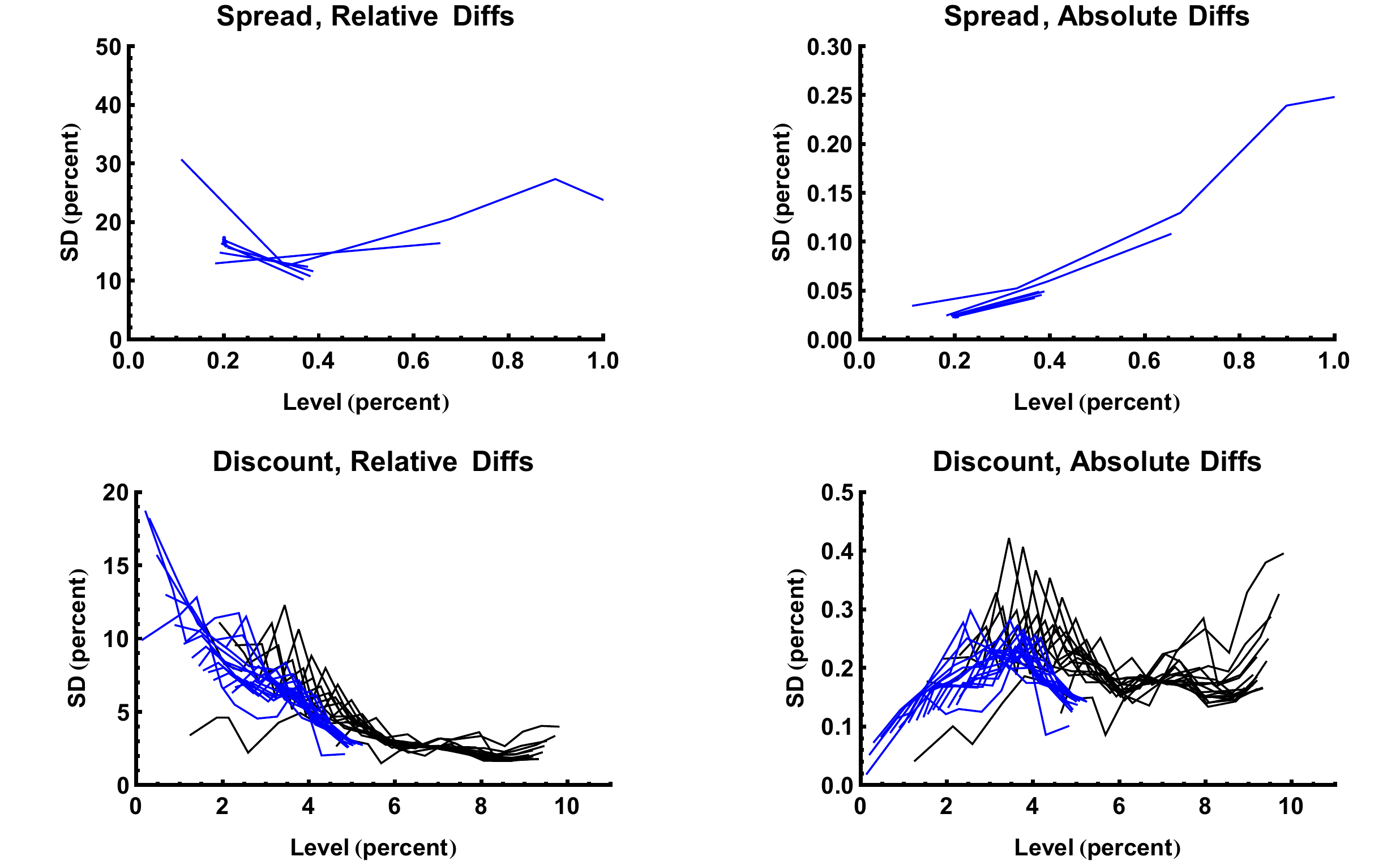}
	\caption{Standard deviation of relative and absolute 10-day differences for spread and discount tenor points on USD interest rate curves.  Tenors used are \{1/4,\ 2,\ \ldots,\ 10,\ 12,\ 15,\ 20,\ 25,\ 30\}-years.  Top panels spread (i.e. projection minus discount), lower panels discount (from OIS).  {\bf Black} data is for 1989-2004 (Libor discounting, no spread) and {\bf Blue} data is for 2004-14 (OIS discounting, and separate spread).}
	\label{f:USDdiffs}
\end{figure}

We analyse 10-day changes of tenor points on the discount and spread curves.  Discounting and projection curves were bootstrapped from OIS, 3M deposits and Libor swaps out to 30 years.  Futures were not included for simplicity and because these do not have fixed tenors (they are relative to calendar points).  Figure \ref{f:USDdiffs} shows standard deviations of relative differences and absolute differences versus level.  The interest rate range was divided into 25 basis point (bp) sections and the standard deviation of the differences at each level calculated.  The level was determined as the median of the levels that started in each 25bps bracket. 

We use standard deviation as our scale metric because it is sensitive to outliers and we are interested in the behaviour of the tails because we are studying VAR and ES.  A robust scale metric, e.g. Median Deviation about the Median, would therefore be inappropriate. 
From Figure \ref{f:USDdiffs} we make two tentative conclusions.
\begin{itemize}
	\item Spread 2004-2014 (Projection minus Discounting), versus level is consistent with a log-Normal model because the curves in the top-left panel (relative differences) are roughly flat relative to the range.  This determination is consistent with the top-right panel (absolute differences) which shows roughly linear curves with a possible intercept around zero (single linear fit:  gradient 95\%\ confidence interval includes zero).
	\item Discount 1989--2014, versus level is consistent with a quadratic-level-dependent log-Normal model because each tenor line on the bottom-left panel is curved and they are relatively close together (single quadratic fit: p-values for both linear and quadratic terms better than 1e-30).  The bottom-right panel (absolute differences) supports this determination because it shows no clear clear pattern.  
	
	Note that the 1989-2004 data (black lines) and 2004-2014 data (blue lines) do not show gaps where they overlap.  This is consistent with market discounting behaviour being as we have assumed (i.e. single-curve up to 2004 and multi-curve after 2007 with mixed, but insignificant effects, behaviour 2004-2007).
\end{itemize}

\paragraph{Stressed VAR(99\%) and stressed ES(97.5\%)}

\begin{table}
\begin{tabular}{cc|cc|cc|cc}
     &    & \multicolumn{2}{c|}{relative} 
				 &  \multicolumn{2}{c|}{$\frac{\text{level-relative}}{\text{relative}}$ }
				 & \multicolumn{2}{c}{$\frac{\text{absolute}}{\text{relative}}$} \\  
		&		& \multicolumn{2}{c|}{\%\ of notional} 
				 &  \multicolumn{2}{c|}{ratio}
				 & \multicolumn{2}{c}{ratio} \\ 
Window & Maturity & SVAR(1\%) & SVAR(99\%) & SVAR(1\%) & SVAR(99\%) & SVAR(1\%) & SVAR(99\%) \\ \hline
 2008-9 & 1 & -0.12 & 0.08 & 63. & 115. & 636. & 592. \\
 2008-9 &  2 & -0.36 & 0.21 & 121. & 143. & 267. & 526. \\
  2008-9 & 5 & -2.16 & 1.67 & 102. & 165. & 144. & 190. \\
  2008-9 & 10 & -5.61 & 5.68 & 93. & 152. & 107. & 157. \\
 2008-9 &  20 & -10.19 & 13.08 & 86. & 133. & 91. & 123. \\
 2008-9 &  30 & -14.24 & 18.99 & 81. & 134. & 84. & 119. \\ \hline
2011-12 & 1 & -0.09 & 0.04 & 106. & 224. & 105. & 161. \\
2011-12 &  2 & -0.47 & 0.42 & 90. & 109. & 101. & 88. \\
2011-12 &  5 & -1.76 & 2.27 & 95. & 101. & 105. & 96. \\
2011-12 &  10 & -3.59 & 4.61 & 99. & 114. & 86. & 113. \\
2011-12 &  20 & -5.79 & 8.14 & 102. & 118. & 81. & 115. \\
2011-12 &  30 & -7.73 & 11.17 & 107. & 121. & 77. & 113. \\
\end{tabular}
\caption{Data-Model dependence of 10-day VAR for USD interest rate swaps for the Financial Crisis window (2008-9) and the Euro-Sovereign Crisis window.  Data Models for differences are: relative; level-relative (i.e. standard deviation scales inversely with level, see text and Figure \ref{f:USDdiffs}); and absolute.}
\label{t:vares}
\end{table}
Table \ref{t:vares} shows SVAR(99\%) for swaps of maturity 1-year to 30-year from two 1-year windows (2008-9 and 2011-12), both calibrated for use at the start of 2014.  The 2008-9 window is for a bank whose single stress period is the Financial crisis.  The 2011-12 window is for a bank whose single stress window is the Euro-Sovereign crisis.  There are  differences because the USD swap market was not stressed during the latter window.  However this is secondary, from the point of view of a single institution, to the major differences due to the Data Models.

\paragraph{Financial crisis window}
\begin{itemize}
	\item With a Data Model of relative differences We can observe that 10-day VAR on a 1-year IRS is roughly 10bps of notional upfront (depending on swap direction), whereas this is 14\%\ or 19\%\ of notional upfront for a 30-year swap.
	\item A level-relative Data Model has VAR up to 60\%\ higher (for 5-year swap) or 40\%\ lower (1-year swap) than a Data Model that uses relative differences.
	\item A Data Model using absolute differences shows much larger differences at the short end ($>$600\% of relative difference).  This is not surprising since absolute differences will be highly significant at the short end since rates have dropped so much relative to the crisis period selected using high USD-SU-CDS spreads.  There is less difference at the long end, roughly $\pm$20\%.  However intermediate maturities have significant differences, up to 50\%\ up to 10 year maturities, relative to a Data Model using relative differences.
	\item Although we do not show ES(97.5\%) for reasons of space, it is very close to VAR(99\%) using this cleaned interest rate data.  The range of differences is roughly $\pm$10\%.  This confirms that {\it for cleaned data} the ES(97.5\%) calibration is appropriate with respect to VAR(99\%).  However, see below for data cleaning effects --- practically speaking this similarity may be purely because of the data cleaning.
\end{itemize}

\paragraph{Euro-Sovereign crisis window}
\begin{itemize}
	\item Apart from the 1-year maturity, all the Data Models agree to within 25\%\ on SVAR(99\%).  This is purely because the stress window is at the same rates level as the current market.  This is the best case for agreement between the Data Models and only applies to {\it spot SVAR and ES}.  Lifetime SVAR and ES will have significantly different results because of scaling differences between the Data Models for higher rates levels.  Agreement on spot SVAR and ES is no predictor of agreement on lifetime values.  
\end{itemize}

\paragraph{Data Cleaning and Extended Rates Levels}

We now analyse the daily Effective Fed Funds rates 1972--2014 to get an idea of a wider rates range (up to 20\%) and compare with interest rates 1989--2014 from swaps data (discount rates).  

\begin{figure}[htbp]
	\centering
		\includegraphics[width=1.1\textwidth,trim=60 0 0 0]{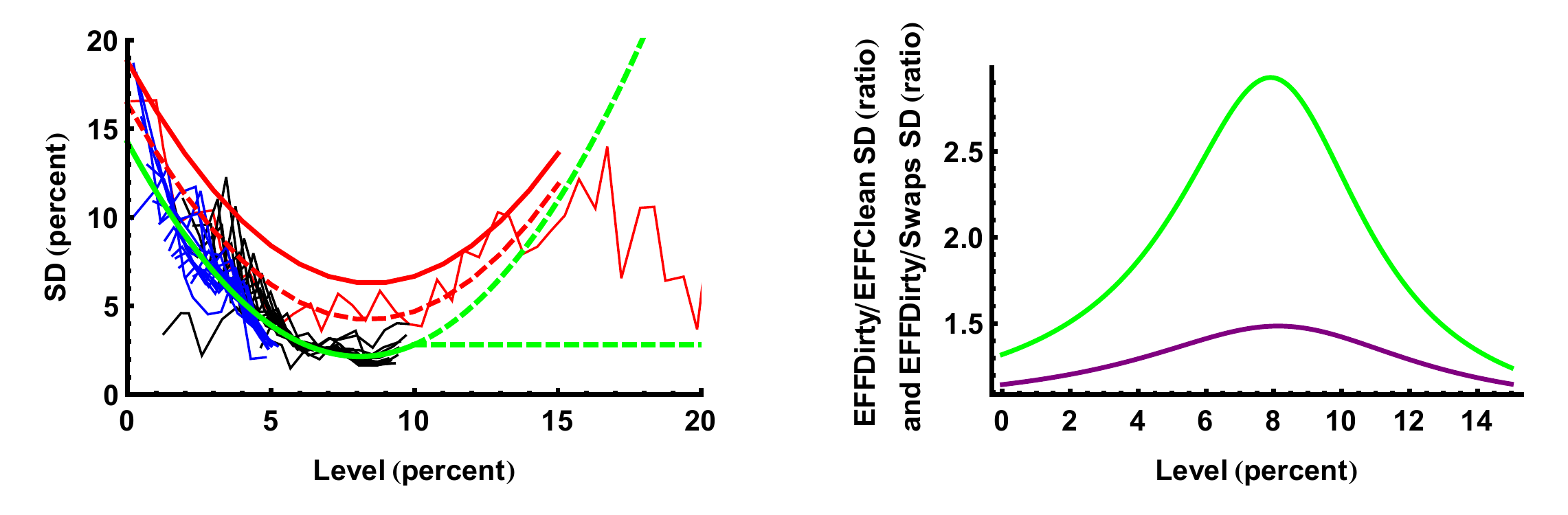}
	\caption{{\bf LHS\ } Standard deviation of relative 10-day differences discount tenor points on USD interest rate curves (black = swaps/Libor data 1989--2004, blue = swaps/OIS data 2004--2014) compared with standard deviation of relative 10-day differences for cleaned Effective Fed Funds rates (red, data from 1972--2014).  {\bf Solid red} curve is quadratic fit to not-cleaned Effective Fed Funds (EFF) data from 0\% to 15\%.  {\bf Dashed  red}  curve is similar quadratic fit to cleaned Effective Fed Funds data.  {\bf Solid green} curve is quadratic fit to cleaned swaps/Libor and swaps/OIS data together.  {\bf Dashed green} curves are two possible extrapolations: quadratic and flat. {\bf RHS purple\ } Ratio of not-cleaned EFF Standard Deviation vs level to cleaned EFF Standard Deviation vs level.  {\bf RHS green\ } Ratio of not-cleaned EFF Standard Deviation vs level to cleaned swaps/Libor and swaps/OIS Standard Deviation vs level.}
	\label{f:range}
\end{figure}

Figure \ref{f:range} (LHS) shows 10-day relative difference for Effective Fed Funds (EFF) rates and USD discount rates together with interpolating quadratic curves and extrapolating curves (quadratic and flat).  The RHS plot shows the relative standard deviations between not-cleaned and cleaned EFF rates and between not-cleaned EFF rates and discount rates from swaps using ratios of the quadratic fits.
\begin{itemize}
	\item A quadratic curve is a good fit (p-values for linear and quadratic terms better than 1e-10) for all three data sets from level of 0\%\ to level of 15\% : not-cleaned EFF (not shown), fit is solid red curve; cleaned EFF (shown as red line), fit is dashed red curve; and cleaned swaps discount data (black and blue data), fit is dashed green curve.
	\item The quadratic fits to the data (SD vs level) are significantly different: maximum differences more than 40\%\ to 250\%\ around levels of 6\%\ to 8\%.
	\item Extrapolating flat from the discount data from swaps gives vastly different results from a quadratic extrapolation.
	\item Beyond the 15\%\ level the EFF standard deviation goes {\it down}.  It does not fit a quadratic curve nor does it fit a flat extrapolation.  This could be because of lack of data or a genuine change in behaviour.
\end{itemize}

Cleaning Fed Funds data consists of removing isolated points, i.e. a point that jumps up and then down on the next day (to within 10\%\ of the previous value).  Degrees of cleaning are possible by considering a jump that goes up-then-down over more than one day.  The data was cleaned for up to five-day up-then-down moves.  There are an infinity of possible cleaning procedures, we use this as a a bound on the effects of cleaning.

The difference between the EFF quadratic fits and the discount fits could be because of tenor, although the discount data covers 3M to 30Y consistently (although the embedded credit and liquidity risk is always 3M).  Alternatively it could be because EFF simply represents a different market.  It could also be because there is some systematically different effect of the EFF cleaning versus the cleaning of the swaps data.  These two are different because with swaps there is curve information as well as temporal information.  Undoubtedly readers will be able to add to this list.  In any case it suggests caution about combining data sources for quantitative use, even from the same currency.

\paragraph{Data Cleaning versus VAR(90\%) and ES(97.5\%)}

\begin{figure}[htbp]
	\centering
		\includegraphics[width=0.9\textwidth,trim=0 0 0 0]{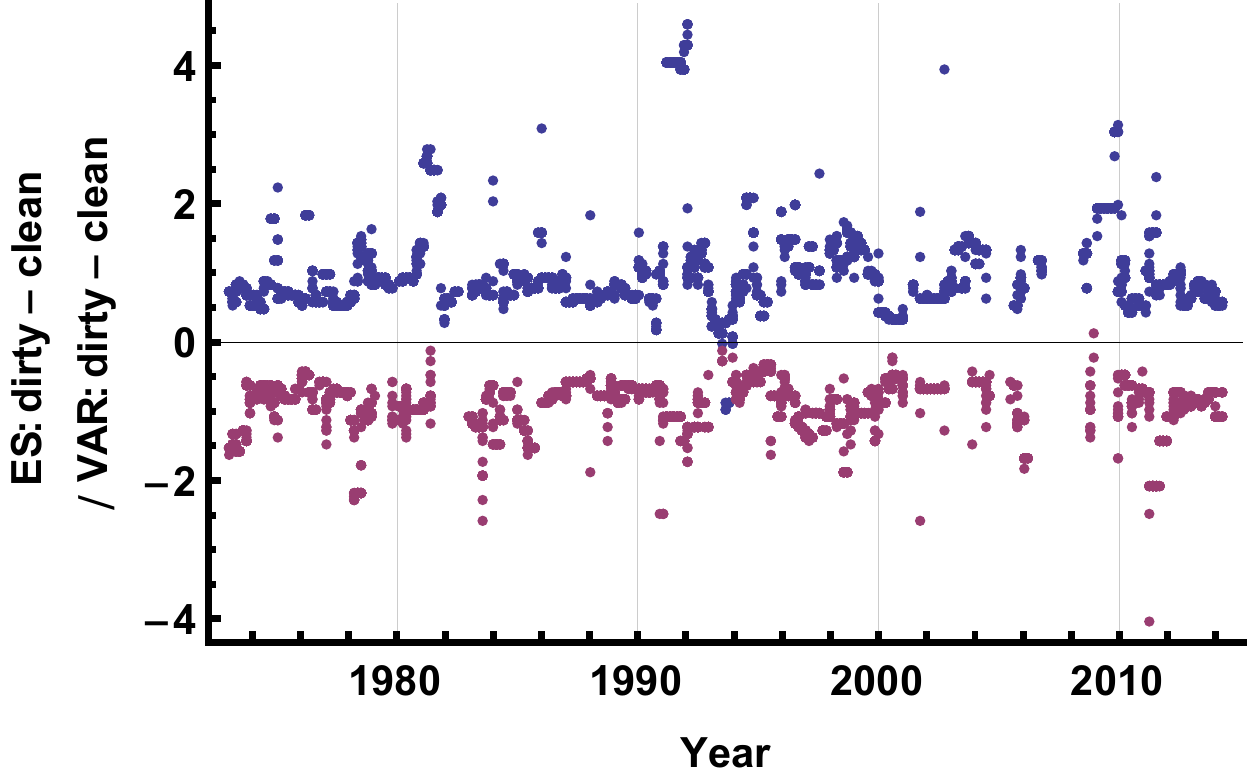}
	\caption{Relative effect of data cleaning on VAR(99\%) and ES(97.5\%) for daily Effective Fed Funds rates.  Running one-year window.  When there is a significant difference between VAR for clean and not-cleaned data we look at the relative change of ES (to avoid dividing by zero).}
	\label{f:cleaning}
\end{figure}
Figure \ref{f:cleaning} show the relative effects of data cleaning on VAR(99\%) and ES(97.5\%) for daily Effective Fed Funds rates.  It uses a one-year rolling window for the VAR and ES computation.  Roughly 70\%\ of the time the ratio is less than one, showing where ES is less affected by data cleaning than VAR.  However, on 30\%\ of the dates ES is much more affected by the data cleaning than VAR.  In fact on those dates it is common to see ES twice as sensitive to the cleaning versus VAR.

For rates levels above 10\%\ we may only have stressed data, e.g. from Oil Shock or 1980s Inflation shocks.  Any model we consider must make some assumption for the discount rates Data Model above 10\%.  We could argue for a flat extrapolation from the levels around up to 10\%.  Alternatively we could argue for quadratic extrapolation as suggested by EFF data.  This may seem immaterial today, however, for lifetime capital costs it is quite possible to get scenarios with high rates levels.  These scenarios affect today's pricing for collateralized interest rate swaps where there is initial margin \cite{Green2014a}, i.e. when traded through Central Counterparties, or for lifetime capital hedging \cite{Green2014b}.   These prices in turn affect collateral calls --- including those by the Central Counterparty \cite{Kenyon2013e}.     Alternatively we could argue that whenever future rates levels are above 10\%\ in simulations then we are in a stressed situation so we do not need historical data.  
  
The number of alternatives, and their materiality, prompts the following Proposals.

\section{Proposals}

The authors in their personal capacity (i.e. with no other representation) make the following proposals for improvement of VAR and ES tail risk metrics.
\begin{enumerate}
	\item Standard data: clean, complete, and common.  That is, all banks (or other users) use identical data.  This can be regulated in the same way that regulations like Basel III are administered.  In fact, given the range of results we have demonstrated with typical choices we would argue that standard data should always be part of capital and funding regulations.
	\item Common Data Models.  That is, all banks (or other users) use identical Data Models.   The required coverage should be limited to where there is reasonable data, and a lookup table approach adopted otherwise.  The prospective Prudent Valuation can be included by specifying a set of Data Models, as it requires \cite{EBA-CP-2013-28}.
	\item Lookup tables for VAR and ES by tenor based on underlying level for USD rates.  For other cases the procedure shown here should be used to identify the key relationships.  The prospective Prudent Valuation can be included by specifying a set of lookup tables (depending on use).  Several tables are required because different banks are sensitive (as in stressed VAR) to different periods.
	 
	\end{enumerate}

Given that we are looking at tail events over a short period, generally 1-year, and used for a {\it much} shorter period (generally 10 days) we are reluctant to propose any principal component based analysis because the percentage of movements described by a fixed number of factors usually decreases with window length.  Additionally,  {\it by definition} the stressed 1-year window is not representative of non-stressed conditions.  Put simply, we do not see the necessity of another level of modeling, or any added value.

These standardisation proposals may seem to create model risk.  This is not the case because we follow the prospective Prudential Valuation in proposing using more than one model.  We would also propose that all models be supported by evidence and derived transparently.  This external standardization also avoids any potential conflict of interest between capital derived from Risk Departments and their business impact within a bank.  

These standardization proposals offer significant efficiency in terms of complexity because all banks would be required to follow the same set of Data Models and lookup tables.  

These proposals would require significantly more effort, and potentially new quantitative skills, from regulators.  However, we have demonstrated that the alternatives are quantitatively highly diverse.  

In short, the problem with tail metrics is not the choice of metric, but the data cleaning and the Data Models, {\em before} the data gets to the metric.

We add a final proposal to remove the pro-cyclicality in current risk calculations using sums of unstressed and stressed results (see Appendix 2):
\begin{itemize}
	\item Replace the sum of unstressed and stressed results by twice the maximum of the two.  This preserves the quantitative level at the height of the next crisis and removes the procyclicality that is otherwise driven by the cyclicality of the non-stressed results.
\end{itemize}

\FloatBarrier
\section{Discussion and Conclusions}

Contrary to usual perception, we have demonstrated that historical VAR and ES, and their stressed counterparts, are not objective measures of tail risk because of their critical dependence on data cleaning and Data Models.  Data cleaning and Data Models are subjective choices.  These tail metrics are not even immutable, any change of data cleaning, or --- especially for CDS --- coverage, by data providers makes comparisons problematic.  

To provide consistency of historical VAR and ES we propose standardization of data (i.e. clean, complete, and common) and standardisation of Data Models.  These must be included in capital and funding regulations because they have such quantitatively significant effects.  Where there is insufficient data or a requirement for simplicity we propose sets of look-up tables adapted to the different windows banks must use according to the regulations described in Section \ref{s:reg}.  By common data and Data Models we mean identical for all users.  To avoid model risk we follow the prospective Prudential Valuation \cite{EBA-CP-2013-28} in proposing using a  set of models, derived transparently from the standard data. 

Model-based VAR and ES provide no advantage with respect to data cleaning and Data Models because these create the input distribution that model-based approaches must calibrate to. 

There are three main determinants of risk tail metrics: data cleaning; Data Model; and choice of VAR or ES.  All three have highly significant effects (50\%\ is not uncommon out 10 years) for USD interest rate swaps from 1-year maturity to 30-year maturity.  We do not expect the results to be qualitatively different for other instruments or currencies because data cleaning and Data Models are initial steps that cannot be avoided in any analysis.  Our data cleaning removed outliers and so with clean data we saw no difference between VAR(99\%) and ES(97.5\%).  VAR(99\%) is already the robust (i.e. median) estimator or losses above VAR(98\%) so the added information from the loss estimation property of ES(97.5\%) is unclear.  It is also unclear whether ES adds anything that is not data-cleaning dependent.  With standard data this might be an area to revisit.

We would also query the need to move to ES for regulatory purposes given that netting sets are legally defined --- it is not clear that the coherence property of ES is relevant here.  Given that ES cannot be used for backtesting its additional value is also questionable.

The numerical results can be queried with respect to details of our data cleaning procedures and choice of Data Models {\em this is precisely the point}.  We have demonstrated that the details of data cleaning and choice of Data Model are critical --- and subjective --- for risk tail metrics.  We have presented what we consider to be reasonable choices, but our choices can certainly be debated.  Hence, a significant quantity of capital is riding on subjective choices.

It can be argued that it is easier to make results look bad, i.e. getting a wide range of quantitative outcomes here, than it is to make results look good, i.e. getting a small range of quantitative outcomes.  However, we have not used models that are outside the literature.  We would also point to the comparative RWA study that found a similarly wide range of outcomes in practice for identical portfolios \cite{BCBS-240}.

Robust statistics is a well-developed field \cite{Huber2009a} and may appear to provide another route forward.  For example median deviation could be used instead of standard deviation.  However, the basic problem for historical VAR/ES is that they have limited effect without additional assumptions on the Data Model.  A further issue is that this would add another level of subjective model choices: why this robust statistic and not another one?  Additionally, we are interested in the tails that are, by definition, not very robust.   For the tail metrics themselves, as a robust proposal we could advocate using the median loss greater than a threshold calibrated to be equal to the current VAR.  This, of course, would be the median loss above 10-day 98\%\ VAR, i.e. 99\%\ VAR.  Thus the current VAR specifications can --- already --- be viewed as robust estimators of lower percentile tail losses.

The tension between tail measures and model specification (not-cleaned data is an example of a mixture model) was previously investigated by \cite{DellAquila2006a} amongst others.  For financial and regulatory purposes their approach of using sophisticated modeling is not appropriate because it loses simplicity and transparency.  In addition the validation problem remains given that we are not only dealing with extremes but with very limited data.  Standardisation is robust, modeling is diverse.

In conclusion we have analysed the data cleaning and Data Model dependencies of VAR and ES, and proposed methods for their resolution.  We proposed standardized data (clean, complete, and common) and standard Data Models; or using lookup tables derived from  transparent analysis of market behaviour based on standard data.  We also proposed that this standard data and the Data Models be part of capital and funding regulations, just as other numerical items are.  The problem with tail risk metrics calculation is not the metrics --- it is everything {\em before} the data gets to the metrics.

\section*{Appendix 1: Data Cleaning for USD Interest Rates}

USD interest rate input data is OIS swaps, 3M deposits, and Libor swaps.  The data is cleaned before bootstrapping into discount and projection curves.  There are two types of cleaning: by date and by instrument.  Since we have a curve of data per date we do each date first, filling in missing data, and then look for false data by instrument.

Data cleaning is ad hoc almost by definition.  We provide this mechanism as one example with no claim to optimality.  On each date.
\begin{itemize}
	\item Interpolate missing instrument values using a monotonic cubic spline \cite{Hyman83}.
	\item Extrapolation:
	\begin{itemize}
	\item If there is a gap of one or two days then linearly interpolate across time.    If this is done then re-apply interpolation as there may now be an interpolation opportunity.
	\item Otherwise extrapolate flat for Libor swaps.  For OIS swaps take the last spread to Libor and extrapolate assuming a constant spread.
	\end{itemize}
\end{itemize}
For instrument data we apply the following cleaning.  This aims to removes isolated points that are bad, i.e. a jump away from previous values followed quickly by a reverse jump.
\begin{itemize}
	\item Check for evidence of bad data.  For each instrument:
	\begin{itemize}
		\item Calculate the standard deviation (SD) of the differences, SD(all).
		\item Remove the biggest $r\%$ of the absolute differences and re-calculate the standard deviation, SD(some), and record the ratio SD(all)/SD(some) = ratio(observed)
		\item Calculate the ratio for a Standard Normal (SN) distribution with the same number of data points, ratio(SN), 256 times.
		\item If ratio(observed) is greater than Mean(ratio(SN))+5 SD(ratio(SN)), then assume there is bad data.
	\end{itemize}
	\item If  detect bad data then:
	\begin{itemize}
	\item Record the $r\%$ quantile of the absolute differences q(r).
	\item Replace any point where it is different from the point before and the point after by q(r), and where the differences either side have different signs.  Replace with the average of the points either side.
	\item Also replace points that jump, stay for one observation, then jump back.
	\end{itemize}
\end{itemize}
We use an $r\%$ such that we would expect it to remove one good point, if there were no bad data points, per couple of thousand data points. $r\%=3\%$ provides roughly this level of security.  Since we are (mostly) interested in single years (i.e. 260) points this is expected to preserve VAR.  Of course changing even one outlier may alter ES, this is a fragile tail metric.

\section*{Appendix 2: Pro-Cyclicality of MR and CVA VAR Capital}

MR and CVA VAR capital use the sum of VAR and SVAR.  Thus at the peak of the next crisis, if it is similar to the last one in intensity, both of these capital terms will see 2$\times$SVAR.  Now if we assume that the current SVAR capital is significant, then as the next crisis unfolds banks will need significant extra capital, in fact they will need exactly as much as they were considered to be missing in the last crisis.  Thus, although the banks are safer from a default point of view their capital problem is unchanged.  This does not seem to be a desired objective of effective capital regulation.

Now suppose that the next crisis is sufficiently far away that banks have fully implemented their counter-cyclical capital buffers (CCB).  If we assume that the CCB are the same size as SVAR-VAR effects, then the combination of VAR+SVAR and CCB is cycle-neutral.  However, the same effect could be achieved by using the maximum of VAR and SVAR with no CCB.  In any case VAR+SVAR reduces the effectiveness CCB, so we would question the utility of having both.

The simplest solution is to have maximum of SVAR and VAR with CCB adjusted to take into account the reduction in opposition to counter-cyclicality that a cycle-neutral MR and CVA VAR produce.

\bibliographystyle{chicago}
\bibliography{kenyon_general}

\end{document}